%% file: paper.tex
\documentclass{svproc}

\input{headers}

\begin{document}

\title{Lightweight Parallel Foundations:\\a model-compliant communication layer}
\titlerunning{Lightweight Parallel Foundations}

\author{Wijnand Suijlen \and A.~N.~Yzelman}
\authorrunning{Suijlen and Yzelman}

\tocauthor{Wijnand Suijlen, A.~N.~Yzelman}

\institute{Paris Research Center,
Huawei Technologies\\
20 quai du Point du Jour,
92100 Boulogne-Billancourt, France\\
\email{\{wijnand.suijlen,albertjan.yzelman\}@huawei.com}}

\maketitle

\begin{abstract}
\input{abstract}
\end{abstract}

\input{content}

\bibliographystyle{spmpsci}
\bibliography{references}

\end{document}

%% file: headers.tex
\usepackage{a4wide}

\usepackage{url}

\usepackage{cite}
\usepackage{amsmath,amssymb}
\usepackage{algorithmic}
\usepackage{graphicx}
\usepackage[caption=false]{subfig}
\usepackage{textcomp}
\usepackage{xcolor}
\usepackage[ruled,linesnumbered]{algorithm2e}
\usepackage{bbm}
\usepackage{todonotes}

\usepackage{listings} 
\newcommand{\bflog}{\textbf{log} \ }

\newcommand{\bsproot}{\texttt{LPF\_ROOT}}

\newcommand{\bigO}{\mathcal{O}}
\newcommand{\bspexec}{\texttt{lpf\_\-exec}}
\newcommand{\bsphook}{\texttt{lpf\_\-hook}}
\newcommand{\bsprehook}{\texttt{lpf\_\-rehook}}
\newcommand{\bspregglob}{\texttt{lpf\_\-register\_\-global}}
\newcommand{\bspregloc}{\texttt{lpf\_\-register\_\-local}}
\newcommand{\bspdereg}{\texttt{lpf\_\-deregister}}
\newcommand{\bspput}{\texttt{lpf\_\-put}}
\newcommand{\bspget}{\texttt{lpf\_\-get}}
\newcommand{\bspsync}{\texttt{lpf\_\-sync}}

\newcommand{\bspresizememreg}{\texttt{lpf\_\-resize\_\-memory\_\-register}}
\newcommand{\bspresizemsgq}{\texttt{lpf\_\-resize\_\-message\_\-queue}}
\newcommand{\bspprobe}{\texttt{lpf\_\-probe}}

\newcommand{\bspargst}{\texttt{lpf\_args\_t}}
\newcommand{\bspinitt}{\texttt{lpf\_init\_t}}
\newcommand{\PlatformBSP}{LPF}
\newcommand{\memcpy}{\texttt{memcpy}}

\newcommand{\mpiput}{\texttt{MPI\_\-Put}}
\newcommand{\mpiget}{\texttt{MPI\_\-Get}}

\newcommand{\mpiisend}{\texttt{MPI\_\-Isend}}
\newcommand{\mpiirsend}{\texttt{MPI\_\-Irsend}}
\newcommand{\mpiprobe}{\texttt{MPI\_\-Probe}}
\newcommand{\mpirecv}{\texttt{MPI\_\-Recv}}
\newcommand{\mpiirecv}{\texttt{MPI\_\-Irecv}}
\newcommand{\mpiwaitall}{\texttt{MPI\_\-Wait\-all}}

\def\BibTeX{{\rm B\kern-.05em{\sc i\kern-.025em b}\kern-.08em
    T\kern-.1667em\lower.7ex\hbox{E}\kern-.125emX}}

%% file: abstract.tex
We present the Lightweight Parallel Foundations (LPF),
an interoperable and model-compliant communication layer adhering to a strict performance model of parallel computations.
LPF consists of twelve primitives, each with strict performance guarantees,
two of which enable interoperability.

We argue that the principles of interoperability and model compliance suffice for the practical use of \emph{immortal algorithms}:
algorithms that are proven optimal once, and valid forever.
These are ideally also implemented once,
and usable from a wide range of sequential and parallel environments.
This paradigm
is evaluated by implementing an immortal fast Fourier transform (FFT) using LPF,
and compared to state-of-the-art FFT implementations.
We find it performs on par to Intel MKL FFT while consistently outperforming FFTW,
thus showing model compliance can be achieved without sacrificing performance.

Interoperability encourages the propagation of immortal algorithms as widely as possible.
We evaluate this by integrating an LPF PageRank into Spark,
without changing any PageRank nor Spark source codes,
and while requiring only a minimal interface layer.


%% file: content.tex
\section{Introduction}
\label{sec:motivation}



We expect an algorithmic complexity analysis to correspond to the run-time of its implementation;
an algorithm,
once designed,
should transfer to different architectures without invalidating its complexity analysis.
This should be no different for parallel algorithms.
The parallel case differs from the sequential only in communication between processing units:
not only should we account for computational work and memory use,
but also for communication patterns and their sizes.
Valiant and McColl coined the term \emph{immortal algorithms} for provably optimal parallel algorithms that should be re-used as broadly as possible~\cite{Valiant1990,Hill1998}.
A communication layer that adheres to the performance model in which optimality is proven is an obvious and minimal prerequisite for portably realising immortal algorithms,
while wide-spread adoption of immortal algorithms remains impractical if their implementations are not also interoperable.



The core principles of the Light\-weight Par\-al\-lel Foundations (LPF) focus on exactly those prerequisites:
model compliance and interoperability.
Section~\ref{sec:interface} presents the twelve LPF primitives,
discusses their semantics,
attaches performance guarantees to each primitive,
illustrates their use, and
describes how LPF manages interoperability.
Section~\ref{sec:implementation} describes four LPF implementations that indeed attain model compliance and allow for wide-ranging interoperability,
thus allowing for the transportable implementation of immortal algorithms.
Section~\ref{exp:bench} demonstrates that
1) LPF achieves model compliance while retaining high throughput and low latency,
2) an immortal fast Fourier transform (FFT) algorithm implemented on LPF compares favourably to that of established high-performance FFT implementations,
and 3) high-performance algorithms implemented in LPF are transparently usable from the Spark Big Data platform
at no modification to the algorithm nor to Spark and at negligible integration costs.
Section~\ref{sec:conclusion} concludes that adhering to our two core principles suffices for implementing usable immortal algorithms,
while enabling the use of immortal algorithms from any parallel framework.


\section{Interface}
\label{sec:interface}

Table~\ref{tab:asymptotic-costs} lists all twelve LPF primitives.
This section discusses their semantics and their performance guarantees that allow for model compliance.
Finally,
we demonstrate their use to achieve interoperability.

\begin{figure}
\centering
\smallskip
\begin{tabular}{r@{:\ \ }l@{\hspace{.5em}}r@{:\ \ }l@{\ }}
lpf\_exec             & $\mathcal{O}(N g+\ell)$, $\mathcal{O}(\ell)$.     & lpf\_put              & $\Theta(1)$.      \\
lpf\_hook             & $\mathcal{O}(N g+\ell)$, $\Theta(1)$.             & lpf\_get              & $\Theta(1)$.      \\
lpf\_rehook           & $\mathcal{O}(N g+\ell)$, $\Theta(1)$.              & lpf\_sync             & $hg+\ell$.\\
\noalign{\smallskip}
lpf\_register\_local  & $\mathcal{O}(M + N)$, $\Theta(1)$.                & lpf\_deregister       & $\Theta(1)$. \\
lpf\_register\_global & $\mathcal{O}(M + N)$, $\Theta(1)$.                & lpf\_probe            & $\Omega(1)$. \\
\end{tabular}
\smallskip

lpf\_resize\_memory\_register: $\mathcal{O}(N)$.\hspace{2em}lpf\_resize\_message\_queue: $\mathcal{O}(N)$.\\
\smallskip
\caption{%
All \PlatformBSP{} primitives and their asymptotic run-time costs.
$M$ depends on the \PlatformBSP{} state,
$N$ depends on function arguments,
and $p$, $g$, $\ell$ are system constants. 
An alternative cost following a big-Oh guarantee 
indicates that a specialised combination of implementation and target hardware can improve on the given guarantee. 
}
\label{tab:asymptotic-costs}
\end{figure}

\subsection{Semantics}
LPF follows the Single Program, Multiple Data (SPMD) pa\-ra\-digm. The 
user writes an SPMD program as a C function and 
executes it using a set of new parallel processes via \bspexec{}, 
or using an existing set of processes via \bsphook{} or \bsprehook{}. 
The LPF run-time state across those processes form a new \emph{context},
which the parallel function receives via its arguments
together with the total number of processes $p$, a unique process ID $s \in \{0,1,\ldots,p-1\}$,
and an \bspargst{} struct through which arbitrary data may be passed.
The \bspargst{} also allow broadcasting function symbols.
All primitives except \bsphook{} require a context as parameter; in 
sequential environments we provide the \bsproot{} default context.
An active 
context can be temporarily replaced by a pristine one by 
\bsprehook{}, which simplifies writing libraries. All primitives except 
\bspput{} and \bspget{} are blocking. A context that calls 
\bspexec{} or \bsprehook{} hence is put on hold as long as the given SPMD 
function runs, ensuring that a process is active in at 
most one context; i.e., active contexts are always disjoint.
Algorithm~\ref{alg:exec-example} illustrates launching 
a parallel context from a sequential one, passing user arguments to the 
parallel section, and receiving an error code in return.

\begin{algorithm}
\begin{lstlisting}
#include <lpf/core.h>
#include <stdlib.h>
void spmd( lpf_t ctx, lpf_pid_t s, lpf_pid_t p, lpf_args_t args);
enum { OK =  0, ILLEGAL_INPUT = 1 };

int main( int argc, char **argv ) {
  int in[2], out = OK;
  lpf_args_t args = LPF_NO_ARGS; 
  in[0] = atoi(argv[1]); in[1] = atoi(argv[2]); 
  args.input  = &in[0]; args.input_size  = sizeof(in);
  args.output = &out;   args.output_size = sizeof(out);
  lpf_exec( LPF_ROOT, LPF_MAX_P, spmd, args );
  return out;
}
\end{lstlisting}
\caption{Example use of \bspexec{}}
\label{alg:exec-example}
\end{algorithm}

The \bspput{} and \bspget{} copy local data to a remote process and vice versa, but do not synchronise.
Both primitives use \emph{memory slots}, offsets, and a size to identify local and remote memory areas.
Memory slots are created by \bspregloc{} if it is only referred to locally or by the collective \bspregglob{}
otherwise.
The \bspdereg{} cancels any registration.
Memory that is the target or source of communication may not be used by non-LPF statements.
LPF primitives are allowed to read from or write to the same memory; the latter is resolved
in some sequential order akin to arbitrary-order CRCW PRAM~\cite{Borodin1985}.
Reading \emph{and} writing to the same memory,
however,
is illegal.
Only a fence by \bspsync{} guarantees completion of communication.
Attributes to \bspsync{}, \bspget{}, and \bspput{} allow LPF extensions 
to relax guarantees for improved performance;
an \bspsync{} attribute assuring absence of write-conflicts, for example,
could improve throughput,
resulting in a lower effective $g$.
 
Algorithm~\ref{alg:hello-world} illustrates how a parallel matrix computation could be bootstrapped.
It retrieves a global input matrix size,
computes the local matrix size and checks if the local dimensions make sense.
Errors, if any, are broadcast and written to output before the function returns.
Note that by exploiting CRCW conflict resolution no buffer is required for error checking.
Although not shown for brevity,
all LPF primitives also return error codes:
either success,
fatal,
or a user-mitigable error.
Errors of the latter type, such as out-of-memory,
will not have side effects.
LPF only maintains local error states, as maintaining a global state requires costly periodical inter-process interactions.
Only \bspsync{}, \bspexec{}, \bsphook{}, and \bsprehook{} may fatally fail due to remote errors,
at the latest when attempting to communicate with an aborted LPF process.
Users, when encountering an error, thus either mitigate locally or clean up and exit the SPMD function;
errors then propagate naturally without causing deadlocks.

\begin{algorithm}
\begin{lstlisting}
void spmd( lpf_t ctx, lpf_pid_t s, lpf_pid_t p, lpf_args_t args ) {
  lpf_memslot_t s_lerr, s_gerr, s_mdim; int M, N, mdim[2];
  lpf_pid_t k, root_pid = 0; size_t offset = 0;
  /* local and global error states */
  int lerr = OK, gerr = OK;

  /* get input */
  mdim[0] = 0; mdim[1] = 0;
  if( args.input_size != 0 )
    memcpy( mdim, args.input, args.input_size );

  /* allocate and activate LPF buffers */
  lpf_resize_memory_register( ctx, 3 );
  lpf_resize_message_queue( ctx, 2*p );
  lpf_sync( ctx, LPF_SYNC_DEFAULT );

  /* register memory areas for communication */
  lpf_register_local(  ctx, &lerr, sizeof(lerr), &s_lerr );
  lpf_register_global( ctx, &gerr, sizeof(gerr), &s_gerr );
  lpf_register_global( ctx, &mdim, sizeof(mdim), &s_mdim );

  /* get global matrix size if we do not have it */
  if( args.input_size == 0 )
    lpf_get(ctx, root_pid, s_mdim, offset, s_mdim, offset,
      sizeof(mdim), LPF_MSG_DEFAULT);
  lpf_sync( ctx, LPF_SYNC_DEFAULT );

  /* compute local matrix size */
  M = (mdim[0]+(int)(p-s-1)) / (int)p;
  N = mdim[1];
  if(p > INT_MAX || M <= 0 || N <= 0) lerr = ILLEGAL_INPUT;

  /* broadcast errors using write-conflict resolution */
  if( lerr != OK ) for( k = 0; k < p; ++k )
    lpf_put( ctx, s_lerr, offset, k, s_gerr, offset,
      sizeof(int), LPF_MSG_DEFAULT );
  lpf_sync( ctx, LPF_SYNC_DEFAULT );

  if( gerr == OK ) { /* build matrix, do compute, ... */ }

  /* clean up & write back error code */
  lpf_deregister( ctx, s_lerr );
  lpf_deregister( ctx, s_gerr );
  lpf_deregister( ctx, s_mdim );
  if( args.output_size == sizeof(int) ) 
    memcpy( args.output, &gerr, args.output_size );
}
\end{lstlisting}
\caption{A `hello world' LPF example}
\label{alg:hello-world}
\end{algorithm}

\subsection{Model-compliance}
We enable model compliance by 1) defining asymptotic complexity guarantees on run-time costs for each primitive in Table~\ref{tab:asymptotic-costs},
while 2) define the maximum time spent on inter-process communication adheres to the BSP model~\cite{Valiant1990}.
The first enables traditional algorithm complexity analysis for process-local programs, while the second enables immortal algorithm design to account for inter-process communication.

Let process $s$ send $t_s$ words and receive $r_s$ words;
in LPF,
these variables can only be modified by use of \bspput{} and \bspget{},
and are reset after each call to \bspsync{} which executes the thus-defined $h$-relation,
with $h=\max_s\{\max\{t_s,r_s\}\}$.
Measured from the time the last process calls \bspsync{},
LPF requires the collective time spent in communication to be, on average, $hg+\ell$ or less,
regardless of the precise pattern or any message or synchronisation attribute used.

Combinations of hardware and software properties may allow for improvement of some Big-Oh guarantees given in Table~\ref{tab:asymptotic-costs};
processes could be spawned in the same address space, for example,
or memory registration could proceed in $\Theta(1)$ time when willing to accept a penalty to $g$ and $\ell$.
The \bspprobe{} is required for implementing immortal algorithms since optimality usually requires parametrisation in $p$, $g$, and $\ell$.
Offline benchmarks such as in Section~\ref{exp:bench} enable implementations to use a $\Theta(1)$ table lookup;
alternatively, we
allow online benchmarks with arbitrary complexity.

LPF also defines strict bounds on its memory use.
The \bspresizememreg{} controls how many memory slots may be registered,
while the \bspresizemsgq{} controls how many RDMA requests the current process can queue or be subject to.
Buffer sizes become active after a fence provided each call completed successfully.
Highly scalable implementations must reserve heap memory linear in the number of reserved memory slots and messages;
smaller parallel systems,
however,
may be allowed an extra $\bigO(p)$ memory in order to improve latency or throughput for implementations on small parallel systems.

\subsection{Interoperability}
An immortal algorithm should integrate trivially with any user application,
be it from a photo editor plug-in, from a data analytics tool on Spark, or from any other application on any sequential or parallel framework.
In LPF,
integrating any SPMD function into a sequential application is a one-step procedure: a call to \bspexec{} with \bsproot{}. 
Integration into an arbitrary parallel environment consists of two steps:
1) initialisation of the library through a platform dependent call which returns an \bspinitt{} object, and
2) one or more calls to \bsphook{} using this \bspinitt{} object.
Algorithm~\ref{alg:hook-example} shows an example of
integrating an LPF function from within a set of $p$ processes connected through an TCP/IP network; 
see Section~\ref{sec:spark} for its application to Spark.
While Algorithm~\ref{alg:hook-example} shows its use when using our MPI-based LPF implementation,
this mechanism is also available for our POSIX threads-based implementation, our ibverbs implementation (which uses MPI for process management), and
(by extension) our hybrid implementations.
The user code can make an arbitrary number of calls to \bsphook{} as long as the given \bspinitt{} remains valid.

\begin{algorithm}
\begin{lstlisting}
#include <lpf/mpi.h>
...
void spmd( lpf_t ctx, lpf_pid_t s, lpf_pid_t p, lpf_args_t args);
...
int main( int argc, char**argv ) {
    char * hostname = NULL, * portname = NULL;
    lpf_pid_t process_id = 0, nprocs = 0;
    lpf_init_t init = LPF_INIT_NONE;
    ... // user code, which also decides on master's hostname,
    ... // master's port name, process_id, and nprocs.
    lpf_mpi_initialize_over_tcp(
        hostname, portname, 30000, //server info and 30-second time-out
        process_id, nprocs,        //process info
        &init
    ); 
    ... //define and initialise args
    lpf_hook( init, &spmd, args );
    ... //continue user code using LPF output
    lpf_mpi_finalize( init );
    return 0;
}
\end{lstlisting}
\caption{Example interoperable use of immortal algorithms via \bsphook{}.}
\label{alg:hook-example}
\end{algorithm}



\section{Implementation}
\label{sec:implementation}

An LPF implementation must remain model-compliant on any hardware platform;
no communication or memory usage pattern may cause run-time guarantees to slip.
Resolving the \bspput{} on a cache-coherent shared-memory system as an array lookup for the memory slot followed by a \memcpy{},
for example,
leads to a non-conforming LPF implementation:
several threads may write into the same cache line,
thus inducing false sharing and a slowdown of up to a factor $p$ during the subsequent \bspsync{}.

We show how to attain the performance guarantees that LPF defines on contemporary parallel architectures by developing three implementations,
targeting
cache-coherent shared memory, 
distributed memory with RDMA communication, and
distributed memory with message-passing systems.
Additionally,
we present a hybrid LPF for clusters of networked multi-core computers that combines the shared-memory implementation with a distributed-memory one.
All distributed-memory implementations allow creation of an \bspinitt{} instance via TCP/IP where one peer must be selected as master.


Our common implementation strategy delays execution of all communication
requests
until the \bspsync{}, so that memory registrations 
and buffer reallocations can be finished before data communication starts.
As summarised by Table~\ref{tab:impl-sync}
the \bspsync{} runs through four phases: 1) a global barrier and 
a first meta-data exchange informing the data destination of each \bspput{}/\bspget{};
2) a write-conflict resolution on the data destination followed 
by a second meta-data exchange informing the data sources what data can be sent 
without overlap; 3) the actual data exchange; 4) a final barrier
that ensures all communication has finished before returning to the user program.

We detail model-compliance aspects of our implementations by
first considering the asymptotic complexities of the algorithms we employ,
and finally by considering the compliance of the underlying infrastructure.

\begin{table}
\centering
\begin{tabular}{ll|l|l|l|l|l||l|l}
            \multicolumn{7}{c||}{Progress in \bspsync{} $\rightarrow$} & \multicolumn{2}{l}{System constants}\\
\multicolumn{2}{l|}{Impl.}& Barr.       &  Meta-data    & Write-conflict  & Data-exch.       & Barr.  & $g$ & $\ell$ \\
\hline
Shared-   &T& $\log p$    & $\mathbf{p}$           &  \multicolumn{2}{|c|}{ $\mathbf{m + h_s + h_b}$}  & $ \log p $   & $1$ & $p$   \\
memory    &M& $1$         & $p$           &  \multicolumn{2}{|c|}{ $m$    }  & $1$        &&\\
	      &A& hierar.     & array   & \multicolumn{2}{|c|}{ dest. thread calls \memcpy{}}  & hierar. &&\\
\hline
RDMA      &T& \multicolumn{2}{|c|}{$\mathbf{p + m}$}  & $\mathbf{m + h_s + h_b/s}$ & $\mathbf{m + h_s + h_b}$  & $\log p$   & $1$  &$p$   \\
\emph{Direct} 
          &M& \multicolumn{2}{|c|}{$ m $}     &$m + h_s + R/s$   & $p + m + h_s$    & $1$           && \\
	      &A& \multicolumn{2}{|c|}{Direct}         & radix-sort    & Put              & tree    &&  \\
\hline
Mesg.     &T& \multicolumn{2}{|c|}{$\mathbf{m \ \bflog{p}}$} & $m + h_s + h_b/s$& $m + h_s + h_b$  & $\mathbf{\bflog p}$     &$\log p$&$\log p$\\
\emph{RB} &M& \multicolumn{2}{|c|}{$m$}     & $m + h_s + R/s$ & $m + h_s$         & $1$             && \\
	      &A& \multicolumn{2}{|c|}{RB}      & radix-sort    & Send/Recv        & tree    &&  \\
\hline
Hybrid    &T& $\mathbf{\bflog q}$ &$\mathbf{m \ \bflog{(p/q)}}$  & $\mathbf{q(m + h_s + h_b/s)}$    & $m + h_s + h_b$  & $\mathbf{\bflog p}$  & $q + \log{(p/q)}$  &$\log{p}$  \\
\emph{RB} &M&  $q$        &$m$              & $m + h_s + R/s$   & $p/q + m + h_s$  & $1$                              & &  \\
	      &A&  hierar.    &  RB         & radix-sort    & Put/\memcpy{}    & comb.   &&  \\
\end{tabular}\smallskip

\caption{Overview of \bspsync{} implementations in terms of algorithms (A)
with their big-Oh time (T) and memory (M) complexities. \emph{RB} and \emph{direct} are
abbreviations for the randomised Bruck and direct all-to-all total meta-data exchanges.
Here, $m$ is the maximum number of messages any process is sending or subject to, while $h_s$ and $h_b$ correspond to the h-relation of small and big messages, respectively, where $s$ is the size limitation of a small message.
The maximum amount of memory registered by any process is $R$,
while $p$ is the number of processes and $q$ is the number of processes per node.
The system constant columns for $g$ and $\ell$,
which are dominated by the time complexities in \textbf{bold}, 
summarise how they scale with $p$ and $q$. }
\label{tab:impl-sync}
\end{table}

\subsection{Algorithmic compliance}

Our shared-memory LPF implementation uses POSIX threads with communication
and synchronisation mechanisms similar to MulticoreBSP~for~C~\cite{Yzelman2014},
which maintains a request queue for each thread pair while an \bspsync{} executes all requests at their destination,
protected by two barriers.
We differ by employing an auto-tuned hierarchical barrier (hierar.) which is faster on systems with many cores~\cite{Nishtala2009}.
The RDMA and message-passing implementations use MPI to govern
and coordinate processes. They resolve communication differently from each other, either using native
ibverbs (RDMA write), MPI one-sided communications (\mpiput{}/\mpiget{}), 
or MPI message-passing (\mpiirsend{}-\mpiirecv{}-\mpiwaitall{} or \mpiisend{}-\mpiprobe{}-\mpirecv{}). 
Our hybrid LPF routes
intra-node communication through our shared-memory implementation,
and routes inter-node communication through a given distributed-memory implementation.
Each memory registration is performed twice: on the thread level,
and on the distributed level.
An \bspput{} or \bspget{} locally decides from the remote process ID which memory slot and which \PlatformBSP{} context to use. 

The meta-data exchanges are all-to-all communications,
which can be a bottleneck on high throughput networks.
An all-to-all with \emph{direct} message exchanges requires at least $p$ messages per process.
As an alternative, following the same intuition as Rao et al.~\cite{Rao1995} but now optimising for latency,
we combine the all-to-all index algorithm by Bruck et al.~\cite{Bruck1997} with two-phase randomised routing by Valiant~\cite{Valiant1990} to reduce this, with high probability,
to $2 \log{p}$ messages at the cost of increasing the total payload size by a factor $\bigO(\log{p})$.
A distributed-memory LPF implementation can employ either algorithm,
leading to trade-offs between latency and throughput as detailed in Table~\ref{tab:impl-sync}.
The asymptotic complexities thus derived show our implementations are model compliant.

\subsection{Infrastructure compliance}

\begin{figure}
\centering
\includegraphics[width=.9\textwidth]{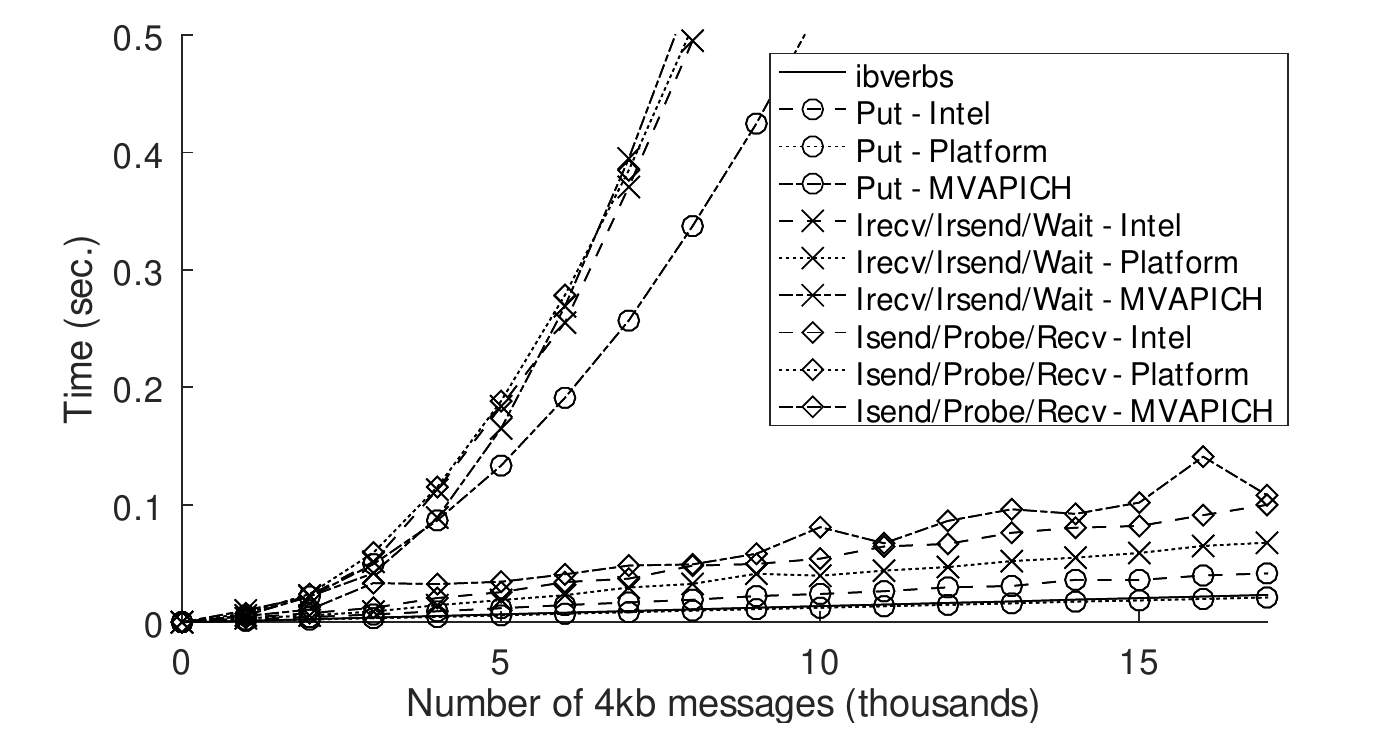}
\caption{Time needed to send $n$ messages round-robin to $p$ processes using
one of the three described methods over an FDR Infiniband network with 4 servers.
A solid line shows the ibverbs baseline performance.
}
\label{fig:FDR-put_and_msg}
\end{figure}

Figure~\ref{fig:FDR-put_and_msg} shows the performance of sending many small $4$~kB messages over an Infiniband FDR network using native ibverbs,
and compares it to the various MPI variants described earlier.
This pattern tests for model compliance (we expect an affine relation),
but also occurs naturally in immortal algorithms
such as the FFT in Section~\ref{sec:fft}, list ranking, or other irregular computations.
The experiment shows an LPF implementation based on MPI must take care in its back-end selection. 
For example,
MPI-based RDMA leads to asymptotic non-compliance when using MVAPICH,
while it leads to model-compliance for IBM Platform MPI.
The ibverbs implementation is consistently compliant,
and is indeed the distributed-memory LPF backend used in subsequent experiments.

For shared-memory architectures,
similar behaviour appears when using MPI back-ends, with some communication methods behaving superlinearly while the pure Pthreads version complies perfectly.

\section{Evaluation}
This section evaluates our model-compliant LPF implementation by
first measuring the system constants $g$ and $\ell$ of our test systems.
Second,
we show the benefit of the design and use of immortal algorithms implemented on a model-compliant communication layer by the comparison of an immortal FFT algorithm versus two state-of-the-art FFT libraries;
experiments show such systematic software design does not hinder performance,
and in fact surpasses Intel MKL FFT performance for mid-size vector sizes.
Third, we showcase how LPF enables interoperability with Spark to show such immortal algorithms can be used in other parallel frameworks.
Experiments employ the systems described in Table~\ref{tab:test-systems} which all run CentOS 7.2.
We use GCC 4.8.5, Intel MPI 2018.2, Intel MKL 2018.2, FFTW 3.3.8, 
Java JRE \& SDK 1.8.0 update 102, Scala 2.11, Hadoop/HDFS 2.7.7, and Spark 2.3.1 where appropriate.

\begin{table}[t]
\centering
\begin{tabular}{r@{\hspace{1em}}l@{\hspace{1em}}l@{\hspace{1em}}l}
Label       & Sandy-$8$       & Ivy-$p$    & BigIvy\\
\hline
\\[\dimexpr-\normalbaselineskip+\smallskipamount]
Huawei server model   & RH2288v2 & RH2288v2 & RH8100v3 \\
Intel CPU   & E5-2650  & E5-2690v2 & E7-8890v2 \\
Cores x sockets x nodes  & $8 \times 2 \times 8$ & $10 \times 2 \times p$ & $15 \times 8 \times 1$  \\
Comp.\  / mem.\ speed per core   & $16$ / $4.3$ & $24$ / $5.1$ & $22.4$ / $2.8$ \\
Network technology & QPI \& FDR ib & QPI \& EDR ib & QPI \\
\hline
\noalign{\smallskip}
\end{tabular}\smallskip

\caption{The test systems. Compute speed is in Gflop/s and memory speed in Gbyte/s.}
\label{tab:test-systems}
\end{table}



\subsection{System constants}
\label{exp:bench}
The BSP model assumes that any $h$-relation can be realised within $T(h) = g h + \ell$ time on average,
which implies that system constants $g$ and $\ell$ can be inferred from the average time
required to fulfil a small number of worst-case communication patterns. 
We focus on the total-exchange up to a maximum volume $n_\text{max}$ 
of at least four times the size of the available cache memory
to ensure we measure out-of-cache behaviour.
We estimate $g$ from $(T(n_\text{max}) - T(2p)) / (n_\text{max} - 2p)$ and $\ell$ from $\max\{ T(0), 2T(p)-T(2p) \}$.
The latter is sensitive to small deviations so we estimate the mean of $T(0), T(p)$ and $T(2p)$ by continuous and random sampling over extended time periods:
Sandy-$8$ and Ivy-$6$ for several days,
and BigIvy for about one month.
These show small 95\%-confidence intervals for $T(h)$, $g$, and $\ell$.
We use a comparable number of cores for each machine and present the results in Table~\ref{tab:bsp-params}.
Our performance on Ivy-$6$ 
is at $90\%$ of capacity when compared against effective intra- and inter-node communication speeds.

\begin{table}
\centering
\begin{tabular}{ll||rl|rl|rl|rl}
\multicolumn{2}{l||}{$1$ word = $w$ bytes}
    & \multicolumn{2}{l|}{$w=8$} & \multicolumn{2}{l|}{$w=64$}&  
                                         \multicolumn{2}{l|}{$w=1024$}& \multicolumn{2}{l}{$w=1048576$} \\
\hline
\hline
&&&&&&&&&\\[\dimexpr-\normalbaselineskip+\smallskipamount]
 \emph{Sandy-8}& $r$ (ns/byte)    & $1.18$ &              & $0.874$ &           & $0.864$&             & $0.777$&              \\
 \emph{Hybrid-RB} & $g$ ($\times$)   & $332$  & $\pm 0.39$   & $82.8$ & $\pm 0.15$ & $22.4$ & $\pm 0.23$  & $6.83$ & $\pm 0.14$   \\
 128 procs    & $\ell$ (words)   & $ 5877$& $\pm 351$    & $ 725$ & $\pm 3.9$  & $ 54 $ & $\pm	0.28$& $0.06$ & $\pm 0.0005$ \\
\hline
&&&&&&&&&\\[\dimexpr-\normalbaselineskip+\smallskipamount]
 \emph{Ivy-6}  & $r$ (ns/byte)    & $0.806$ &             & $0.730$&            & $0.719$&             & $0.653$&\\
 \emph{Hybrid-RB} & $g$ ($\times$)   & $303$  & $\pm	0.11$   &$80.8$  & $\pm	0.046$& $13.5$ & $\pm  0.056$& $2.75$ & $\pm	0.01$  \\
 120 procs   & $\ell$ (words)   & $7717$ & $\pm	178$    & $706$  & $\pm	5.2$  & $179$  & $\pm	31$  & $0.06$ & $\pm 0.0003$ \\
 \hline
&&&&&&&&&\\[\dimexpr-\normalbaselineskip+\smallskipamount]
 \emph{BigIvy}   & $r$ (ns/byte)   & $0.844$&             & $0.806$ &             & $0.769$&            & $0.825$ \\
 \emph{Pthreads} & $g$ ($\times$)   & $51.9$ & $\pm	0.26$  & $10.7$  & $\pm 0.060$ & $5.63$ & $\pm 0.041$& $5.43$ & $\pm 0.52$  \\
 120 procs       &  $\ell$ (words)& $6231$ & $\pm	74$    & $1086$  & $\pm	11$    & $100$  & $\pm 0.93$ & $4.3$  & $\pm 3.2$  
\end{tabular} \smallskip
\caption{The system constants $g, \ell$ normalised w.r.t. $r$, the speed of a \memcpy{}. 
The unit of communication is $w$ bytes.
The $\pm$ indicate the size of a 95\% confidence interval.}
\label{tab:bsp-params}
\end{table}

%

\subsection{Performance}
\label{sec:fft}
To show we achieve model compliance without impeding performance,
this section compares the state-of-the-art in parallel FFT libraries versus a known immortal FFT algorithm by Bisseling and Inda~\cite{Valiant1990,Inda2001}.
We use its HPBSP implementation~\cite{Yzelman2014},
which runs on LPF by use of an BSPlib layer on top of LPF;
this layer enables the use of a large body of BSP algorithms originally written for BSPlib.
Being able to implement such complete higher-level libraries additionally demonstrates the expressiveness of LPF.
The HPBSP FFT relies on either FFTW or Spiral for process-local FFTs,
of which there are $n/p^2$ before and $p$ after a global data redistribution,
where $n$ is the vector size and $\sqrt{n}>p$.
We modified the code to use the Intel MKL FFT instead.
Since none of FFTW, Spiral, and MKL expose unordered time-shifted FFTs 
our implementation must manually twiddle and permute after redistribution,
resulting in two additional passes over the output vector compared to what would have been optimal.

\begin{figure}
	\centerline{%
		\includegraphics[trim= 20 5 20 5, clip, width=0.5\textwidth]{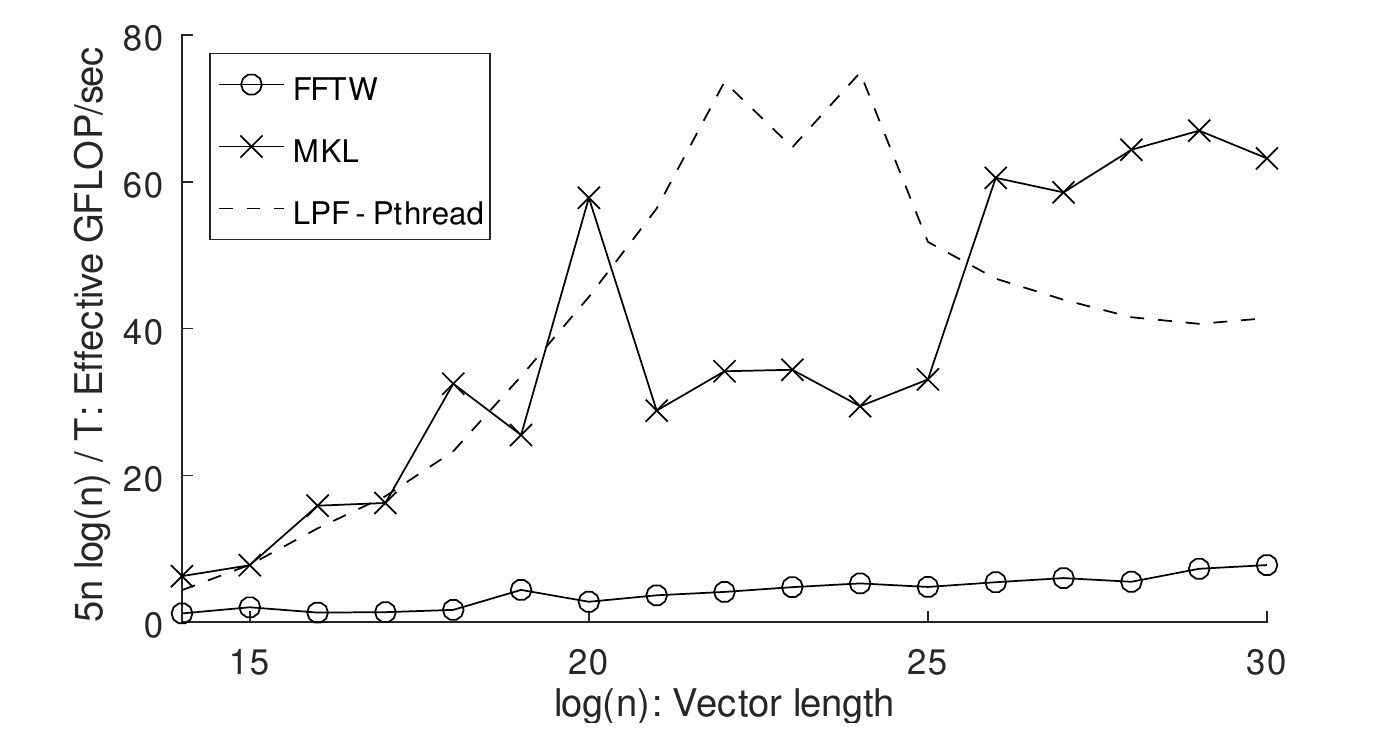}%
		\includegraphics[trim= 20 5 30 5, clip,width=0.5\textwidth]{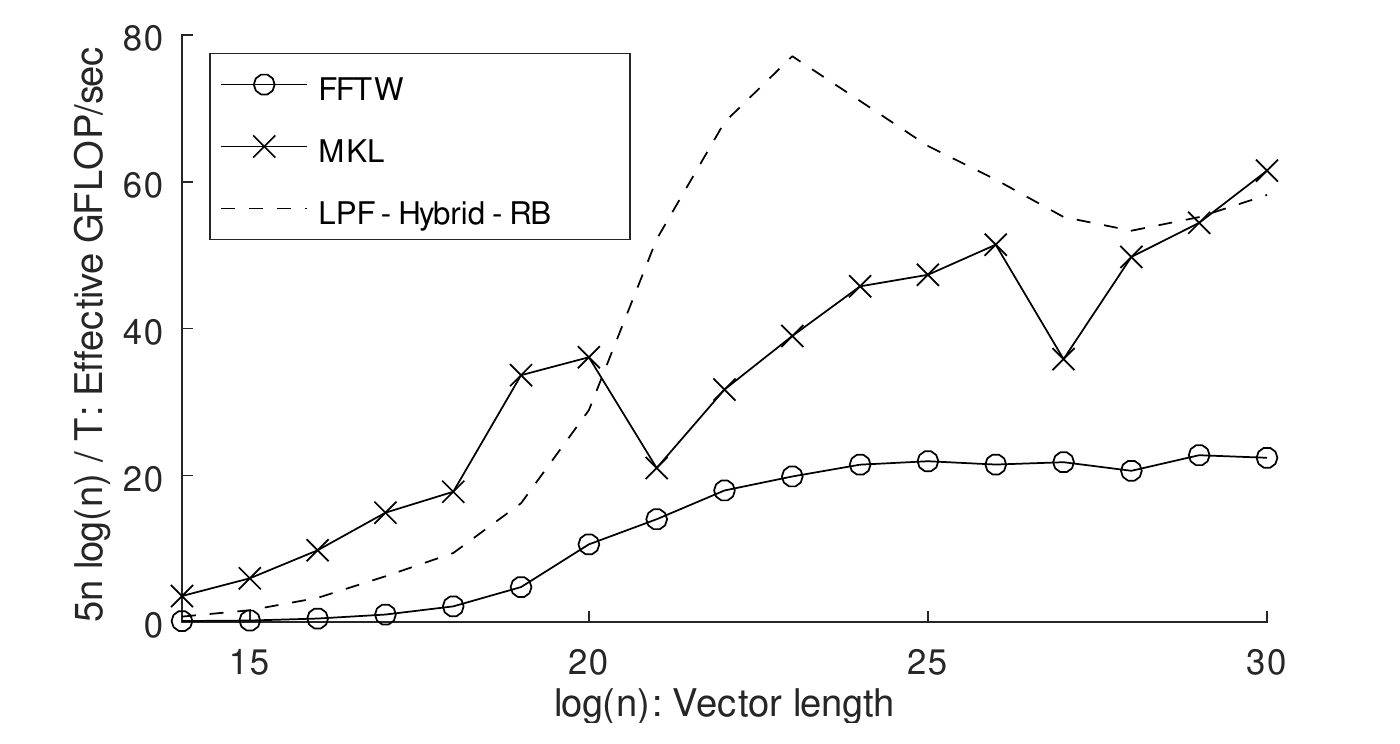}}
	\caption{HPBSP compared with FFTW3 and MKL on BigIvy (left) and Sandy-8 (right).}
	\label{fig:FFT}
\end{figure}

We perform experiments for vectors lengths $n=2^k$, $14\leq k \leq 30$.
For each $N$ we perform $200$ transforms and record the average time taken in Figure~\ref{fig:FFT},
using the Pthreads LPF implementation on BigIvy (left) and the hybrid implementation on Sandy-8 (right).
For BigIvy we compare against the multi-threaded FFTW and MKL FFTs,
where we repeated the experiments for 64, 128, and (when possible) 120 threads,
retaining only the fastest results.
For Sandy-8 we compare the MPI implementations using $128$ processes only.
HPBSP is faster than FFTW for both platforms and performs on par with MKL.
On the distributed-machine, we observing speedups up to 2.5x for vector lengths between $2^{20}$ and $2^{29}$ versus MKL with differences becoming negligible for larger factors, which is normal as growing the problem size makes the computation increasingly compute-bound.
On the shared-memory machine,
we perform similarly for lengths $2^{20}$ and less,
and better when between $2^{21}$--$2^{25}$.
The immortal FFT implemented on LPF is outperformed by MKL for larger vectors on this large shared-memory machine;
one cause may be the extra twiddles our implementation incurs due to the unavailability of highly-optimised time-shifted FFT implementations,
causing a $2$x slowdown as vectors are streamed once more than necessary.
This cost may be hidden from the distributed-memory results since it does not affect inter-process communication.

Despite strict model compliance and despite using an extra level of abstraction (BSPlib),
these experiments show an immortal BSP algorithm still compares favourably in terms of raw performance to state-of-the-art alternatives on this communication-bound algorithm.


\subsection{Interoperability}
\label{sec:spark}
We demonstrate interoperability by calling a high-per\-for\-mance LPF-based PageRank from Spark.
We implement one by directly translating its canonical linear algebra formulation (see, e.g., Langville~\cite{langville2011}) into GraphBLAS,
for which we have a hybrid LPF/OpenMP C++ implementation.
Compared to executing any other native code in Spark,
clients must first collect the worker's hostnames, remove all duplicates, and broadcast them as an array.
Then,
each time before executing LPF code,
each worker derives $p$, $s$, and a master node from the broadcast array.
These variables suffice for creating a valid \bspinitt{} using the TCP/IP mechanism of our distributed LPF implementations,
which enables calling our PageRank (or any other LPF algorithm) via \bsphook{}, any number of times.
All steps from the creation of a \bspinitt{} object must be executed through JNI,
while all prior operations can be implemented in pure Spark and Scala.

Table~\ref{tab:spark} compares running the LPF PageRank versus a typical pure Spark implementation
which does not handle contributions from dangling nodes nor checks for convergence.
While the LPF PageRank does implement these,
this Spark version seems canonical\footnote{Such as~\url{https://github.com/apache/spark/blob/v2.3.1/examples/src/main/scala/org/apache/spark/examples/SparkPageRank.scala}.} and can only skew our comparison in favour of Spark;
hence we compare as is.
We use the
cage15, uk-2002, and clueweb12, 
matrices from SuiteSparse and WebGraph,
all in uncompressed MatrixMarket format.
Pure Spark used 1500 RDD partitions for cage15 and 4500 for uk-2002,
and checkpoints every ten iterations to break lineages and prevent out-of-memory errors.
It did not reach $n_\epsilon$ on uk-2002 within four hours,
and could not complete one iteration for clueweb12 due to out-of-memory errors.

For $n=1$ we mainly measure I/O and start-up costs.
The difference is due to Spark loading the matrix from HDFS while our GraphBLAS employs parallel I/O;
the LPF PageRank could also ingest the matrix data directly from Spark just as any other JNI code,
but this would preclude experiments using clueweb12 due to out-of-memory errors.
We observe speedups of several orders of magnitude in terms of end-to-end performance for larger $n$,
and even greater speedups when comparing time-per-iteration.
By use of LPF interoperability,
Spark is furthermore able to tackle significantly larger problems on the same limited number of nodes,
at very little effort,
and in a way that easily extends to other Big Data platforms.

\begin{table}[t]
\centering
\smallskip
\begin{tabular}{r@{\hspace{1em}}r@{\ \ }r@{\ \ }r@{\hspace{1.5em}}rrrrr@{\hspace{1.5em}}rrr@{\hspace{.5em}}r}
           &      &         &            & \multicolumn{4}{c}{Pure Spark}                 &   & \multicolumn{4}{c}{Accelerated Spark} \\
           & GB   & Gnz    &$n_\epsilon$& $n=1$   & $n=10$   & $n=n_\epsilon $ & $s/$it.  &   & $n=1$     & $n=10$  & $n=n_\epsilon$ & $s/$it. \\
\noalign{\smallskip}
cage15     & $2.5$& $0.1$   &$56$        & $39.6$  & $42.7$   & $1296.9$        &  $22.8$ &   & $5.5$     & $7.1$   & $19.2$         & $0.25$ \\

uk-2002    & $4.7$ & $0.3$ &$73$        & $168.6$ & $1373.8$ & \raisebox{1pt}{\scalebox{0.77}{$>$}}$4$ hrs
                                                                               &$133.9$   &   & $8.7$     & $13.9$  & $48.7$         & $0.56$ \\

clueweb12  &$786$ & $42.5$ &$45$        & -       & -        & -               & -        &   & $658.8$   & $963.2$ & $1875.0$       & $27.7$ \\
\end{tabular}\smallskip

\caption{Pure vs.\ LPF PageRank using Spark on Ivy-$10$,
in seconds,
for $n$ iterations. The matrix sizes are shown in Gbyte and billions of nonzeroes.
LPF requires $n_\epsilon$ iterations to reach a $\epsilon = 10^{-7}$ tolerance.
The seconds-per-iteration is computed by subtracting the time from $n=n_\epsilon$ from that of $n=1$, and dividing the result by $n_e-1$,
except for pure Spark on uk-2002 where $n=10$ is used instead.
The result is rounded down for pure Spark and rounded up for accelerated Spark.}
\label{tab:spark}
\end{table}

\section{Related Work}

We categorise related communication layers by their core algorithmic model;
those based on CSP (message passing) versus those based on PRAM (shared memory and remote direct memory access, RDMA).
MPI 
provides both and is the standard for high-performance message passing.
It refrains from performance guarantees to not impose too large a burden on developers~\cite{Traff2010}.
Indeed, many good implementations exist and support many different systems.
While some argue a usable degree of performance portability is possible without guarantees~\cite{Traff2010},
others have reported that commonly expected asymptotic bounds did not hold in practice~\cite{Balaji2008,Flajslik2016}.

BSPlib~\cite{Hill1998}, UPC~\cite{Carlson1999}, OpenSHMEM~\cite{Chapman2010}, and ibverbs all expose PRAM-style communication,
either by assuming a shared memory or by providing explicit RDMA primitives.
\PlatformBSP{} is rooted in Bulk Synchronous Parallel (BSP)~\cite{Valiant1990} and is much alike BSPlib,
which,
in turn,
was influenced by Cray SHMEM.
While BSPlib implied performance guarantees conform the BSP model,
we define explicit ones for each primitive in terms of time, memory, and communication.
In comparison,
\PlatformBSP{} lacks buffered RDMA and one-sided message passing
while BSPlib lacks primitives for transparent interoperability and introspection of 
performance guarantees.
We also improve error handling, structuring of parallel 
sections, composability of subprograms, and RDMA memory registration. 

In Section~\ref{sec:spark} we call \PlatformBSP{} code from Spark to demonstrate interoperability.
Similarly,
Alchemist~\cite{Gittens:2018:ALD:3219819.3219927} offloads Spark computations to MPI via TCP sockets.
It employs a server-worker architecture
which must be disjoint from the Spark deployment.
We differ from Alchemist by repurposing the Spark worker processes as LPF processes,
which allows direct access to input and output data.
Our approach extends to any parallel framework beyond Spark without requiring any change to LPF nor to its algorithms.

\section{Conclusions \& Future Work}
\label{sec:conclusion}

We present \PlatformBSP{},
a vehicle for implementing portable immortal algorithms.
It does so by adhering to two core principles: model-compliance and interoperability.
Compliance consists of defining asymptotic performance guarantees,
and defining absolute costs for inter-process communication.
Easy-to-use interoperability is designed to help the practical adoption of immortal algorithms.
Requiring only TCP/IP connection and a master node selection,
LPF indeed achieves interoperability with a wide range of parallel and big data frameworks.

We also showed that LPF
establishes good encapsulation of sub-algorithms using \bsprehook{},
allows precise run-time control of message buffers to cope with significant memory constraints, and
includes efficient write-conflict resolution.
We implicitly demonstrated LPF is expressive enough to implement more user-friendly higher-level interfaces by during experiments having made use of an LPF-based collectives library, an LPF BSPlib interface, and a hybrid GraphBLAS interface.

LPF is light\-weight since we achieve our core principles without sacrificing performance,
and foundational since it allows for the implementation of immortal algorithms while facilitating their use as broadly as possible.
Our LPF implementations achieve good performance,
as demonstrated by an immortal FFT implementation that performs on par with established counterparts from Intel MKL and FFTW.
We furthermore applied our interoperability mechanism to run unaltered LPF codes within Spark while requiring only minimal bootstrapping,
and without requiring modifications to Spark source codes.
These show the two core concepts of LPF are indeed sufficient for the implementation and practical application of immortal algorithms.

\subsection*{Future Work}

For future work we foresee extensions via RDMA and synchronisation attributes,
these could enable stale-synchronicity~\cite{xing15},
zero-cost synchronisation~\cite{alpert97},
lower effective $g$ by assuring the LPF implementation there are no overlapping writes,
and more.
While we did not expand on the formal semantics of each LPF primitives in this text,
these additionally allow for proofs of correctness of parallel codes,
and allow for auto-parallelisation.

\medskip{}
\textbf{Acknowledgements} The authors would like to thank 
Ga\'etan Hains, Arvid Jakobsson, Pierre Leca, Bill McColl, Jonathan Nash, Daniel Di Nardo, Antoine Petitet, Dimitris Vekris, and Xia Yinglong
for their insights, feedback, and early adoption.